\magnification1200
\centerline
{\bf  Universal derivation of the asymptotic charges of bosonic massless particles }
\vskip 1cm
\centerline{\bf Kevin Nguyen${}^\dagger$  and  Peter West${}^\ddagger$ }
\vskip 1.2cm
\centerline{ {${}^\dagger {}^\ddagger$\it Department of Mathematics, King's College, London WC2R 2LS, UK }}
\vskip 1.2cm
\centerline{{ ${}^\ddagger$\it Mathematical Institute, University of Oxford, Woodstock Road, Oxford, OX2 6GG, UK}}
\vskip 1.7cm
\centerline{\sl Abstract}
We present a unified treatment of the conserved asymptotic charges associated with any bosonic massless particle in any spacetime dimension. In particular we provide master formulae for the asymptotic charges and the central extensions in the corresponding charge algebras. These formulae can be explicitly evaluated for any given theory. For illustration we apply them to electromagnetism and gravity, thereby recovering earlier results.
\noindent
\vskip 6cm
emails: kevin.nguyen@kcl.ac.uk, peter.west540@gmail.com
\vfill

\eject
In this letter we will derive the conserved asymptotic charges at spatial infinity for any massless particle of integer spin in D spacetime dimensions. Such a derivation was sketched in reference [1] but here we give a complete derivation. An extensive list of references to related subjects on asymptotic charges in given in  reference [1], while the literature on the general theory with broad applicability is relatively scarce [2-8]. Charges associated to Killing symmetries of higher-spin fields have been studied in [6]. However Killing symmetries only form a subset of the asymptotic symmetries to be discussed here as the latter act nontrivially on the gauge fields. A free particle is by definition an irreducible representation of the Poincare group [9]. A Lorentz covariant formulation can be found by embedding this representation into a representation of the Lorentz group carried by a field denoted $A_{\bullet}$. Since this is a larger representation it must satisfy projection conditions. These  are the on-shell conditions which are equivalent to the free equations of motion for the particle [10-16]. It is known that the free equation of motion for any particle can be written in terms of an equation of two spacetime derivatives acting on the field  $A_{\bullet}$ [17-19]. The subscript $\bullet $ represents the Lorentz indices that the field carries. The action for the field  $A_{\bullet}$ then has the form
$$
S=-{1\over 2}\int d^D x\, \sqrt {-\det g}\, \nabla_a A_{\bullet } N^{a b{\bullet } {\bullet }^\prime} \nabla_b A_{\bullet ^\prime}  \,,
\eqno(1)$$
Although the particle is travelling in Minkowski space we will use curvilinear coordinates $x^\mu=(t, r, \theta^m)$ where $\theta^m$ are any coordinates on the sphere $S^{D-2}$ of radius $r$. In particular the asymptotic charges are integrals over the sphere at arbitrary time $t$ and infinite radius limit $r \to \infty$. The most common coordinate choice would be the spherical coordinates $x^\mu=(t,r,\theta, \varphi)$ in four spacetime dimensions. The symbol $\nabla_a$ is the tangent space covariant derivative with respect to the flat metric $\eta_{ab}$, while $g_{\mu\nu}$ is the flat metric in curvilinear coordinates $x^\mu$.  It is given in terms of the vierbein on the sphere by the usual expression $g_{mn}=\hat e_n {}^\alpha \delta_{\alpha \beta} \hat e_m{}^\beta$.  As a result $\det g=-r^{D-2}\det \gamma=-r^{D-2}(\det \hat e )^2$. We refer the reader to reference [1] for further details. The constants $N^{a b{\bullet } {\bullet }^\prime}$ specify how the indices are contracted and obey the relation
$$N^{0 b{\bullet } {\bullet }^\prime}=N^{b0 {\bullet }^\prime {\bullet } } ,
\eqno(2)$$
 which follows from the symmetrical way the fields $A_{\bullet}$ appears in the action.
\par
The variation of the action has the form
$$
\delta S=\int d^D x\, \delta A_{\bullet }  N^{a b{\bullet } {\bullet }^\prime}( \nabla_a  \nabla_b A_{\bullet^\prime } ) \sqrt {-\det g}-\int d^D x\,  \partial _a (\delta A_{\bullet }  N^{a b{\bullet } {\bullet }^\prime}\nabla_b A_{\bullet^\prime }  \sqrt {-\det g} )\,,
\eqno(3)$$
and we read off the equation of motion to be  given by
$$
N^{a b{\bullet } {\bullet }^\prime} \nabla _a\nabla_b  A_{\bullet^\prime}=0\,.
\eqno(4)$$
Since we are dealing with massless particles we typically have a gauge symmetry which we may write as
$$
\delta A_{\bullet } = M^{c}{}_{ \bullet}{}^{ \star} \nabla_c \Lambda_{ \star}\,,
\eqno(5)$$
where $\star$ denotes the indices carried by the gauge parameter $\Lambda_\star$. The gauge invariance of the equation of motion  requires that
$$
N^{a b{\bullet } {\bullet }^\prime} M^{c}{}_{ \bullet^\prime}{}^{ \star} \nabla _a\nabla_b\nabla_c \Lambda_{ \star}=0\,.
\eqno(6)$$
Since the gauge parameter $\Lambda_{ \star}$ is an arbitrary function we require
$$
N^{(a b |{\bullet } {\bullet }^\prime} M^{|c)}{}_{ \bullet^\prime}{}^{ \star} =0\,.
\eqno(7)$$
\par
As an illustration, we give the constants $N^{a b{\bullet } {\bullet }^\prime}$ and $M^{c}{}_{ \bullet}{}^{ \star}$ for Maxwell (spin 1) and Fierz-Pauli (spin 2) theories. In Maxwell theory the constants are given by
$$
N^{ab\, c\, d}=\eta^{ab} \eta^{cd}-\eta^{ad}\eta^{bc}\,, \qquad M^a {}_b=\delta^a_b\,,
\eqno(8)$$
while in the spin two Fierz-Pauli theory, that is, gravity they are given by
$$
N^{ab\, cd\, ef}=2 \eta^{ab}  \eta^{cd} \eta^{ef}-\eta^{ab} \eta^{ce}\eta^{df}-\eta^{ab} \eta^{cf} \eta^{de}+\eta^{ac} \eta^{be} \eta^{df}+\eta^{ac}\eta^{bf}\eta^{de}-\eta^{ac}\eta^{bd}\eta^{ef}
\eqno(9)
$$
$$
+\eta^{ad}\eta^{be}\eta^{cf}+\eta^{ad}\eta^{bf}\eta^{ce}-\eta^{ad}\eta^{bc}\eta^{ef}-\eta^{af}\eta^{be}\eta^{cd}-\eta^{ae}\eta^{bf}\eta^{cd}\,,
$$
$$
M^c{}_{ab}{}^d=\delta_a^c \delta_b^d+\delta_a^d \delta_b^c\,.
\eqno(10)$$
It can be explicitly verified that they satisfy all of the above properties.

\par
The boundary term in the variation of the action is be read off  from equation (3) and is given by
$$
-\int dt d\Omega r^{D-2}\, \delta A_{\bullet }  N^{r b{\bullet } {\bullet }^\prime}\nabla_b A_{\bullet^\prime }\,.
\eqno(11)$$
Examining this boundary term and demanding that it vanish as $r\to \infty$ we  find the asymptotic behaviour   for the fields in tangent space must be  as follows
$$
A_{\bullet} = {A^{(1)}_{\bullet}(\theta^m) \over r^{{D-2\over 2}}} +O(r^{-{D\over 2}})\,, \ \  \Lambda_{\star}= {{\Lambda^{(0)}_{\star}(\theta^m) }\over   r^{{(D-4)\over 2}}}+O( r^{-{(D-2)\over 2}})\,.
\eqno(12)$$
The fall off of the gauge parameter is deduced from that of the field taking account of the fact that $\nabla $ increases the fall off by a power of $r^{-1}$. In particular the falloffs of the symmetry parameters are such that the falloffs of the gauge fields are preserved.  This fact and the reason why the leading term does not depend on time are discussed in reference [1].
\par
As explained in reference [1] since the asymptotic charges involve integrals over the sphere at infinity where the fields are weak we can compute them from the free theory. The advantage is that the free theory is rather  simple and so the procedure is very transparent. We will also follow the tangent space approach of reference [1] as this was found to lead to considerable simplification. The symplectic potential is given by
$$
 \theta[\delta A]=  \int dr d^{D-2}\theta\, \delta A_{\bullet }  N^{0 b{\bullet } {\bullet }^\prime} \nabla_b A_{\bullet^\prime }  \sqrt {-\det g}\,.
\eqno(13)$$
The variation of the asymptotic conserved charges under the gauge transformation with parameter $\Lambda$ can be computed   using the symplectic  methods and it is  given by
$$
\delta Q_\Lambda = \delta_\Lambda \theta [\delta A]- \delta \theta [\delta_\Lambda A]\,,
\eqno(14)$$
where $\delta_\Lambda$ is the gauge variation and $\delta$ is any generic variation.
Applying this to the symplectic potential we find that
$$
\delta Q_\Lambda= \int dr d^{D-2}\theta \sqrt {-\det g}\, (N^{0 b{\bullet } {\bullet }^\prime}
M^{c}{}_{ \bullet^\prime}{}^{ \star}  \delta A _\bullet \nabla_b\nabla_c \Lambda_\star
-N^{0b {\bullet } {\bullet }^\prime}M^{c}{}_{ \bullet}{}^{ \star}\nabla_c \Lambda_{\star}  \nabla_b \delta A_{\bullet^\prime})\,.
\eqno(15)$$
Using the relation of equation (2), we may write the charge as
$$
Q_\Lambda= \int dr d^{D-2}\theta \sqrt {-\det g}\, (N^{0 b{\bullet } {\bullet }^\prime}M^{c}{}_{ \bullet^\prime}{}^{ \star} A_{\bullet}\nabla_b\nabla_c \Lambda_\star - N^{b0{\bullet } {\bullet }^\prime}M^{c}{}_{ \bullet^\prime}{}^{ \star} \nabla _ b A_{\bullet}\nabla_c \Lambda_\star)\,.
\eqno(16)$$
\par
Up to now  all the manipulations in the paper have been  rather obvious but we now have to show that the expression for the charge $Q_\Lambda$ of equation (16) can be expressed as  an integral over the sphere at infinity.
We begin by writing the indices $a,b,\ldots $  explicitly in terms $0$ and $j, k, \ldots =1,\ldots, D-1$, whereupon this charge becomes  
$$
Q_\Lambda= \int dr d^{D-2}\theta \sqrt {-\det g}\, \{
N^{0 i{\bullet } {\bullet }^\prime}M^{0}{}_{ \bullet^\prime}{}^{ \star} A_{\bullet}\nabla_i\nabla_0 \Lambda_\star - N^{i0{\bullet } {\bullet }^\prime}M^{0}{}_{ \bullet^\prime}{}^{ \star} \nabla _ i A_{\bullet}\nabla_0 \Lambda_\star
$$
$$
+N^{0 0{\bullet } {\bullet }^\prime}M^{j}{}_{ \bullet^\prime}{}^{ \star} (A_{\bullet}\nabla_0\nabla_j \Lambda_\star - \nabla _ 0 A_{\bullet}\nabla_j \Lambda_\star)
\eqno(17)
$$
$$
+N^{0 i{\bullet } {\bullet }^\prime}M^{j}{}_{ \bullet^\prime}{}^{ \star} A_{\bullet}\nabla_i\nabla_j \Lambda_\star - N^{i0{\bullet } {\bullet }^\prime}M^{j}{}_{ \bullet^\prime}{}^{ \star} \nabla _ i A_{\bullet}\nabla_j \Lambda_\star  \}\,.
$$
In deriving this equation we have used equation (6) and in particular that $N^{00 |{\bullet } {\bullet }^\prime} M^{| 0}{}_{ \bullet^\prime}{}^{ \star} =0$.
\par
Using equation (7), the first, second  and third   terms of equation (17) can be expressed as
$$
\int dr d^{D-2}\theta \sqrt {-\det g}\, \{
N^{0 0{\bullet } {\bullet }^\prime}M^{i}{}_{ \bullet^\prime}{}^{ \star} \nabla _i (A_{\bullet}\nabla_0 \Lambda_\star )
+ N^{0 i{\bullet } {\bullet }^\prime}M^{0}{}_{ \bullet^\prime}{}^{ \star} \nabla _i (A_{\bullet}\nabla_0 \Lambda_\star )\}
$$
$$
=  -\int  d\Omega r^{D-2}  
N^{r 0{\bullet } {\bullet }^\prime}M^{0}{}_{ \bullet^\prime}{}^{ \star} A_{\bullet}\nabla_0 \Lambda_\star\,.
\eqno(18)$$
 The fourth  term can be written as
$$
-\int dr d^{D-2}\theta \sqrt {-\det g}\, N^{0 0{\bullet } {\bullet }^\prime}M^{j}{}_{ \bullet^\prime}{}^{ \star}\{
 \nabla_j(\nabla_0 A_\bullet \Lambda_\star )- \nabla_j\nabla_0 A_\bullet \Lambda_\star\}\,.
\eqno(19)$$
Using the equation of motion (4) we find that the last term in equation (19) can be written as
$$
 \int dr d^{D-2}\theta \sqrt {-\det g}\, N^{ij{\bullet } {\bullet }^\prime}M^{0}{}_{ \bullet^\prime}{}^{ \star}
 \nabla_i\nabla_j A_\bullet \Lambda_\star \,.
 \eqno(20)$$
 The fifth term in equation (17) can be written as
$$
 \int dr d^{D-2}\theta \sqrt {-\det g}\, N^{0 i{\bullet } {\bullet }^\prime}M^{j}{}_{ \bullet^\prime}{}^{ \star}\{
 \nabla_i (A_\bullet \nabla_j\Lambda_\star )- \nabla_j(\nabla_i A_\bullet \Lambda_\star)+ \nabla_i\nabla_j A_\bullet \Lambda_\star \}\,,
 \eqno(21)$$
while the sixth term in equation (17) can be written as
 $$
 \int dr d^{D-2}\theta \sqrt {-\det g}\, N^{i 0\bullet  {\bullet }^\prime}M^{j}{}_{ \bullet^\prime}{}^{ \star}\{
 -\nabla_j (\nabla_i A_\bullet \Lambda_\star )+ \nabla_j\nabla_i A_\bullet \Lambda_\star \}\,.
 \eqno(22)$$
We find that the last terms in equations (21) and (22) cancel the term in equation (20) using equation (4).
 \par
 We are now left with terms that are indeed integrals over a sphere at infinity and collecting these up we find that the conserved charge is given by
 $$
 Q_\Lambda= \int d\Omega r^{D-2}  \{
 - N^{r 0\bullet  {\bullet }^\prime}M^{0}{}_{ \bullet^\prime}{}^{ \star}A_\bullet \nabla_0 \Lambda_\star
- N^{0 0\bullet  {\bullet }^\prime}M^{r}{}_{ \bullet^\prime}{}^{ \star}\nabla_0 A_\bullet  \Lambda_\star
+N^{0 r\bullet  {\bullet }^\prime}M^{j}{}_{ \bullet^\prime}{}^{ \star}A_\bullet \nabla_j \Lambda_\star
$$
$$
-N^{0 j\bullet  {\bullet }^\prime}M^{r}{}_{ \bullet^\prime}{}^{ \star}\nabla_j A_\bullet  \Lambda_\star
-N^{j0\bullet  {\bullet }^\prime}M^{r}{}_{ \bullet^\prime}{}^{ \star}\nabla_j  A_\bullet \Lambda_\star \}\,.
 \eqno(23)$$
This is the master formula for the conserved asymptotic charges.
Using the asymptotic expansion of equation (12) we see that the charge is finite and non-zero. To apply this master formula to any given theory, one simply need to specify the constants $N^{ab \bullet \bullet'}$ and $M^a{}_{\bullet}{}^\star$.  
\par
Evaluating this charge for the case of Maxwell we find that
$$
Q_\Lambda=\int d\Omega\, r^{D-2}\, \Lambda\, (\nabla_0 A_r-\nabla_r A_0)\,,
\eqno(24)$$
which is the usual expression for the radial electric field integrated against the symmetry parameter $\Lambda$.
Evaluating it for gravity we get
$$
Q_{\xi^0}= 2\int d\Omega r^{D-2}(-\nabla^r \xi^0 (h^\mu{}_\mu -2 h^0{}_0) +2 \xi^0\nabla^r h^m{}_m -2 \xi^0\nabla_m h^m{}^r -2 h^r{}_0\nabla^0 \xi^0)\,,
$$
$$
Q_{\xi^r}= 2\int d\Omega  r^{D-2}(\nabla^0 \xi^r (h^\mu{}_\mu -2h^r{}_r) -2 \xi^r\nabla^0 h^m{}_m +2 \xi^r\nabla_m h^m{}^0 +2h^0{}_r\nabla^r\xi^r)\,,
\eqno(25)$$
$$
Q_{\xi^m}= 2\int d\Omega  r^{D-2}(-2 h^r{}_m \nabla^0 \xi^m +2 \xi^m(\nabla^0h_m{}^r-\nabla^r h^0{}_m) +2 h^0{}_m \nabla^r \xi^m )\,,
$$
in terms of the vector field $\xi^\mu$ that parametrises linear diffeomorphisms. These expressions were obtained in [1] without the use of the general formula (23).
\par
The above computation can be applied to any formulation of the massless particle including the different dual formulations where one generally finds different asymptotic charges.
\par
We can compute the Poisson brackets of the charges by computing their symmetry variation, that is, as
$$
\{Q_{\bar \Lambda},Q_{\Lambda} \}=\delta_{\bar \Lambda} Q_{\Lambda}
$$
$$
= \int d\Omega r^{D-2} \{-N^{r0 \bullet \bullet'} M^0{}_{\bullet'}{} ^{\star}M^c{}_{\bullet}{}^{\star^\prime}\, \nabla_c \bar \Lambda_{\star^\prime} \nabla_0 \Lambda _{\star}\,
-N^{0 0 \bullet \bullet'} M^r{}_{\bullet'}{} ^{\star}M^c{}_{\bullet}{}^{\star^\prime}\, \nabla_c   \nabla_0\bar \Lambda_{\star^\prime} \Lambda _{\star}\,
$$
$$
+N^{0r \bullet \bullet'} M^j{}_{\bullet'}{} ^{\star}M^c{}_{\bullet}{}^{\star^\prime}\, \nabla_c \bar \Lambda_{\star^\prime} \nabla_j \Lambda _{\star}\,
\eqno(26)
$$
$$
-N^{0 j \bullet \bullet'} M^r{}_{\bullet'}{} ^{\star} M^c{}_{\bullet}{}^{\star^\prime}\, \nabla_j   \nabla_c\bar \Lambda_{\star^\prime} \Lambda _{\star}\,
-N^{j 0 \bullet \bullet'} M^r{}_{\bullet'}{} ^{\star}M^c{}_{\bullet}{}^{\star^\prime}\, \nabla_j   \nabla_c\bar \Lambda_{\star^\prime} \Lambda _{\star}\}\,.
$$
 In this equation the index $c$ takes the range $c=0,1,\ldots ,D-1$ while $j=1,\ldots , D-1$ as before.
We evaluate this expression  in the appendix to find the result
$$
\{Q_{\bar \Lambda},Q_{\Lambda} \}={1\over 2} \int d\Omega\, r^{D-2}\, C(\bar \Lambda,\Lambda)\,,
\eqno(27)
$$
where
$$
C(\bar \Lambda,\Lambda)=
N^{0r \bullet \bullet' } M^0{}_{\bullet '}{} ^{\star^\prime}M^0{}_\bullet{}^\star\, \nabla_0 \Lambda _{\star^\prime} \nabla_0 \bar \Lambda_\star +N^{0r \bullet \bullet' } M^r{}_{\bullet '}{} ^{\star^\prime}M^r{}_\bullet{}^\star\, \nabla_r \Lambda _{\star^\prime} \nabla_r \bar \Lambda_\star  
$$
$$
+2 N^{0r \bullet \bullet' } M^r{}_{\bullet '}{} ^{\star^\prime}M^0{}_\bullet{}^\star\, \nabla_r \Lambda _{\star^\prime} \nabla_0 \bar \Lambda_\star+2N^{0r \bullet \bullet' } M^m {}_{\bullet '}{} ^{\star^\prime}M^0{}_\bullet{}^\star\, \nabla_m \Lambda _{\star^\prime} \nabla_0 \bar \Lambda_\star
$$
$$
+2N^{0r \bullet \bullet' } M^r {}_{\bullet '}{} ^{\star^\prime}M^m{}_\bullet{}^\star\, \nabla_r \Lambda _{\star^\prime} \nabla_m \bar \Lambda_\star
\eqno(28)
$$
$$
(N^{0r \bullet \bullet'} M^{m}{}_{\bullet'}{} ^{\star^\prime}M^{n}{}_{\bullet}{}^\star+N^{(0m) \bullet \bullet'} M^{r}{}_{\bullet'}{} ^{\star^\prime}M^{n}{}_{\bullet}{}^\star+N^{(0n) \bullet \bullet'} M^{r}{}_{\bullet'}{} ^{\star^\prime}M^{m}{}_{\bullet}{}^\star
$$
$$
+{2 \over 3} N^{(rn) \bullet \bullet'} M^m{}_{\bullet'}{} ^{\star^\prime}M^0{}_{\bullet}{}^\star+{2 \over 3} N^{(0m) \bullet \bullet'} M^n{}_{\bullet'}{} ^{\star^\prime}M^r{}_{\bullet}{}^\star)\, \nabla_m \Lambda _{\star^\prime} \nabla_n \bar \Lambda_\star
$$
$$
- (\Lambda \leftrightarrow \bar \Lambda )\,,
$$
which is explicitly anti-symmetrical in the gauge parameters as it should be. Note that the function $C(\Lambda, \bar \Lambda)$ is field-independent and is therefore interpreted as a central term in the charge algebra. In general the algebra computed form the theory would be non-abelian but these terms would arise from   terms cubic in the fields. It would be an interesting exercise to find these.
\par
 Evaluating the central term for Maxwell theory we find that it vanishes. For the case of gravity we find that it is given by
 $$
C(\xi,\bar \xi)=8( \nabla_0 \xi^0\, \nabla_0 \bar \xi^r+\nabla_r \xi^r\, \nabla_r \bar \xi^0+\nabla_0 \xi^\mu\, \nabla_r \bar \xi_\mu+\nabla_0 \xi^m\, \nabla_m \bar \xi^r+\nabla_m \xi^m\, \nabla_0 \bar \xi^r
$$
$$
+\nabla_r \xi^m\, \nabla_m \bar \xi^0+\nabla_m \xi^m\, \nabla_r \bar \xi^0+\nabla_m\xi^0\,  \nabla^m \bar \xi^r )-(\xi \leftrightarrow \bar \xi)\,.
\eqno(29)$$
When expanded in terms of asymptotic quantites, one explicitly recovers equation (4.14) of reference [1] at leading order in $r \to \infty$.
 
\medskip
{\bf Ackowledgement}
\medskip

We would like to thank the STFC, grant numbers ST/P000258/1   and ST/T000759/1, for support.  

\medskip
{\bf Appendix}
\medskip

In this appendix we give the detailed  derivation of the expression for the central term in the charge algebra given in equation  (27)-(28). The first step is straightforward and  consists in performing a symmetry variation of the charges  of equation (23). The resulting expression can be organised according to the different derivatives that occur and is given by
$$
\delta_{\bar \Lambda} Q_\Lambda={1 \over 2} \int d\Omega\, r^{D-2}\, (C_{00}+C_{0r}+C_{0m}+C_{rr}+C_{rm}+C_{mn} )\,,
\eqno(A.1)$$
Using the identity $N^{00 \bullet \bullet'} M^0{}_{\bullet'}{}^\star=0$ the first term is given by
$$
C_{00}=-2N^{r0 \bullet \bullet'} M^0{}_{\bullet'}{} ^{\star^\prime}M^0{}_{\bullet}{}^\star\, \partial_0 \Lambda _{\star^\prime}\, \partial_0 \bar \Lambda_\star
= (N^{0r \bullet \bullet'}-N^{r0 \bullet \bullet'}) M^0{}_{\bullet'}{} ^{\star^\prime}M^0{}_{\bullet}{}^\star\, \partial_0 \Lambda _{\star^\prime}\, \partial_0 \bar \Lambda_\star\,.
\eqno(A.2)$$
After using the identity of equation (2) and relabelling some indices, the result can be written as
$$
C_{00}=N^{0r \bullet \bullet'} M^0{}_{\bullet'}{} ^{\star^\prime}M^0{}_{\bullet}{}^\star\, (\partial_0 \Lambda _{\star^\prime}\, \partial_0 \bar \Lambda_\star-\partial_0 \bar \Lambda _{\star^\prime}\, \partial_0 \Lambda_\star)\,.
\eqno(A.3)$$
The terms $C_{0r}, C_{0m}$ can be evaluated in  a similar fashion through repeated use of equations (7) and (2) to find an expressions which are also  manifestly anti-symmetrical under the interchange $\Lambda \leftrightarrow \bar \Lambda$.
\par
We now evaluate the remaining terms which are collectively given by
$$
C_{ij}=2N^{0r \bullet \bullet'} M^j{}_{\bullet'}{} ^{\star^\prime}M^i{}_{\bullet}{}^\star\, \nabla_j \Lambda _{\star^\prime}\, \nabla_i \bar \Lambda_\star-4N^{(0i) \bullet \bullet'} M^r{}_{\bullet'}{} ^{\star^\prime}M^j{}_{\bullet}{}^\star\, \Lambda _{\star^\prime}\, \nabla_i \nabla_j  \bar \Lambda_\star\,,
\eqno(A.4)$$
which can be written, after using (7) on the second term, as
$$
C_{ij}=2N^{0r \bullet \bullet'} M^j{}_{\bullet'}{} ^{\star^\prime}M^i{}_{\bullet}{}^\star\, \nabla_j \Lambda _{\star^\prime}\, \nabla_i \bar \Lambda_\star+2N^{(ij) \bullet \bullet'} M^r{}_{\bullet'}{} ^{\star^\prime}M^0{}_{\bullet}{}^\star\, \Lambda _{\star^\prime} \nabla_i \nabla_j  \bar \Lambda_\star\,.
\eqno(A.5)$$
Now we aim at moving one of the derivative in the second term upon integration. For that we first write the integral over the sphere as a three-dimensional integral over space,
$$
\int d\Omega\, r^{D-2}\, N^{(ij) \bullet \bullet'} M^r{}_{\bullet'}{} ^{\star^\prime}M^0{}_{\bullet}{}^\star\, \Lambda _{\star^\prime} \nabla_i \nabla_j  \bar \Lambda_\star
$$
$$
=\int dr d^{D-2} \theta\, \sqrt{-\det g}\, \nabla_k( N^{(ij) \bullet \bullet'} M^k{}_{\bullet'}{} ^{\star^\prime}M^0{}_{\bullet}{}^\star\, \Lambda _{\star^\prime} \nabla_i \nabla_j  \bar \Lambda_\star )
$$
$$
=\int dr d^{D-2}\theta\, \sqrt{-\det g} \, \nabla_i\nabla_k (N^{(ij) \bullet \bullet'} M^k{}_{\bullet'}{} ^{\star^\prime}M^0{}_{\bullet}{}^\star\, \Lambda _{\star^\prime}  \nabla_j  \bar \Lambda_\star)
$$
$$
-\int dr d^{D-2}\theta\, \sqrt{-\det g}\, \nabla_k(N^{(ij) \bullet \bullet'} M^k{}_{\bullet'}{} ^{\star^\prime}M^0{}_{\bullet}{}^\star\, \nabla_i \Lambda _{\star^\prime}  \nabla_j  \bar \Lambda_\star )
$$
$$
=\int dr d^{D-2}\theta \, \sqrt{-\det g}\,  \nabla_i(N^{(ij) \bullet \bullet'} M^k{}_{\bullet'}{} ^{\star^\prime}M^0{}_{\bullet}{}^\star\, \nabla_k (\Lambda _{\star^\prime}  \nabla_j  \bar \Lambda_\star))
$$
$$
-\int dr d^{D-2}\theta\, \sqrt{-\det g}\, \nabla_k(N^{(ij) \bullet \bullet'} M^k{}_{\bullet'}{} ^{\star^\prime}M^0{}_{\bullet}{}^\star\, \nabla_i \Lambda _{\star^\prime}  \nabla_j  \bar \Lambda_\star )
\eqno(A.6)
$$
$$
=\int d\Omega\, r^{D-2} \,( N^{(ri) \bullet \bullet'} M^j{}_{\bullet'}{} ^{\star^\prime}M^0{}_{\bullet}{}^\star-N^{(ij) \bullet \bullet'} M^r{}_{\bullet'}{} ^{\star^\prime}M^0{}_{\bullet}{}^\star) \nabla_j \Lambda _{\star^\prime}  \nabla_i  \bar \Lambda_\star
$$
$$
+\int d\Omega\, r^{D-2}\, N^{(ri) \bullet \bullet'} M^j{}_{\bullet'}{} ^{\star^\prime}M^0{}_{\bullet}{}^\star\, \Lambda _{\star^\prime}  \nabla_i \nabla_j  \bar \Lambda_\star
$$
$$
=-\int d\Omega\, r^{D-2} (N^{(rj) \bullet \bullet'} M^i{}_{\bullet'}{} ^{\star^\prime}M^0{}_{\bullet}{}^\star+2N^{(ij) \bullet \bullet'} M^r{}_{\bullet'}{} ^{\star^\prime}M^0{}_{\bullet}{}^\star) \nabla_j \Lambda _{\star^\prime}  \nabla_i  \bar \Lambda_\star
$$
$$
-{1\over 2} \int d\Omega\, r^{D-2}\, N^{(ij) \bullet \bullet'} M^r{}_{\bullet'}{} ^{\star^\prime}M^0{}_{\bullet}{}^\star\, \Lambda _{\star^\prime}  \nabla_i \nabla_j  \bar \Lambda_\star\,,
$$
where we used the fact that $\nabla_i$ derivatives annihilate the tensors $N$ and $M$ as they are entirely made out of the metric tensor, and where the last equality again follows from application of equation (7) to the first and third term. Thus we infer the useful identity
$$
\int d\Omega\, N^{(ij) \bullet \bullet'} M^r{}_{\bullet'}{} ^{\star^\prime}M^0{}_{\bullet}{}^\star\, \Lambda _{\star^\prime} \nabla_i \nabla_j  \bar \Lambda_\star
$$
$$
=-{2 \over 3} \int d\Omega\,  (N^{(rj) \bullet \bullet'} M^i{}_{\bullet'}{} ^{\star^\prime}M^0{}_{\bullet}{}^\star+2N^{(ij) \bullet \bullet'} M^r{}_{\bullet'}{} ^{\star^\prime}M^0{}_{\bullet}{}^\star) \nabla_j \Lambda _{\star^\prime}  \nabla_i  \bar \Lambda_\star\,.
\eqno(A.7)
$$
Using this result in equation (A.5) we find that
$$
3\, C_{ij}= (6N^{0r \bullet \bullet'} M^j{}_{\bullet'}{} ^{\star^\prime}M^i{}_{\bullet}{}^\star-4N^{(rj) \bullet \bullet'} M^i{}_{\bullet'}{} ^{\star^\prime}M^0{}_{\bullet}{}^\star-8N^{(ij) \bullet \bullet'} M^r{}_{\bullet'}{} ^{\star^\prime}M^0{}_{\bullet}{}^\star) \nabla_j \Lambda _{\star^\prime} \nabla_i \bar \Lambda_\star\,.
\eqno(A.8)$$
Using this expression the terms $C_{rr}, C_{rm}$ and $C_{mn}$ can now be treated separately. The first two are rather straightforward to analyse and we will only further discuss the term $C_{mn}$, explicitly given by
$$
3\, C_{m n}=(6N^{0r \bullet \bullet'} M^m{}_{\bullet'}{} ^{\star^\prime}M^n{}_{\bullet}{}^\star-4N^{(rm) \bullet \bullet'} M^n{}_{\bullet'}{} ^{\star^\prime}M^0{}_{\bullet}{}^\star
$$
$$
-8N^{(m n) \bullet \bullet'} M^r{}_{\bullet'}{} ^{\star^\prime}M^0{}_{\bullet}{}^\star) \nabla_m \Lambda _{\star^\prime} \nabla_n \bar \Lambda_\star\,.
\eqno(A.9)$$
We split it into two groups of terms. The first one is
$$
3 C^{(1)}_{m n}=-(4N^{(rm) \bullet \bullet'} M^n{}_{\bullet'}{} ^{\star^\prime}M^0{}_{\bullet}{}^\star+2N^{(m n) \bullet \bullet'} M^r{}_{\bullet'}{} ^{\star^\prime}M^0{}_{\bullet}{}^\star) \nabla_m \Lambda _{\star^\prime} \nabla_n \bar \Lambda_\star
$$
$$
=(2N^{(rn) \bullet \bullet'} M^m{}_{\bullet'}{} ^{\star^\prime}M^0{}_{\bullet}{}^\star-2N^{(rm) \bullet \bullet'} M^n{}_{\bullet'}{} ^{\star^\prime}M^0{}_{\bullet}{}^\star) \nabla_m \Lambda _{\star^\prime} \nabla_n \bar \Lambda_\star
$$
$$
=(2N^{(rn) \bullet \bullet'} M^m{}_{\bullet'}{} ^{\star^\prime}M^0{}_{\bullet}{}^\star+2N^{(0m) \bullet \bullet'} M^n{}_{\bullet'}{} ^{\star^\prime}M^r{}_{\bullet}{}^\star) (\nabla_m \Lambda _{\star^\prime} \nabla_n \bar \Lambda_\star-\nabla_m \bar \Lambda _{\star^\prime} \nabla_n \Lambda_\star)\,,
\eqno(A.10)$$
while the second one therefore is
$$
C^{(2)}_{m n}=(2N^{0r \bullet \bullet'} M^m{}_{\bullet'}{} ^{\star^\prime}M^n{}_{\bullet}{}^\star-2N^{(m n) \bullet \bullet'} M^r{}_{\bullet'}{} ^{\star^\prime}M^0{}_{\bullet}{}^\star)\, \nabla_m \Lambda _{\star^\prime} \nabla_n \bar \Lambda_\star\,.
\eqno(A.11)$$
Using the identity
$$
N^{(m n) \bullet \bullet'} M^r{}_{\bullet'}{} ^{\star^\prime}M^0{}_{\bullet}{}^\star=N^{(0r) \bullet \bullet'} M^{(m}{}_{\bullet'}{} ^{\star^\prime}M^{n)}{}_{\bullet}{}^\star
$$
$$
-N^{(0m) \bullet \bullet'} M^{[r}{}_{\bullet'}{} ^{\star^\prime}M^{n]}{}_{\bullet}{}^\star-N^{(0n) \bullet \bullet'} M^{[r}{}_{\bullet'}{} ^{\star^\prime}M^{m]}{}_{\bullet}{}^\star
\eqno(A.12)$$
this second group of terms can be written
$$
(N^{0r \bullet \bullet'} M^{m}{}_{\bullet'}{} ^{\star^\prime}M^{n}{}_{\bullet}{}^\star+N^{(0m) \bullet \bullet'} M^{r}{}_{\bullet'}{} ^{\star^\prime}M^{n}{}_{\bullet}{}^\star
$$
$$
+N^{(0n) \bullet \bullet'} M^{r}{}_{\bullet'}{} ^{\star^\prime}M^{m}{}_{\bullet}{}^\star) (\nabla_m \Lambda _{\star^\prime} \nabla_n \bar \Lambda_\star-(\Lambda \leftrightarrow \bar \Lambda))\,.
\eqno(A.13)$$
Hence we conclude
$$
C_{m n}=(N^{0r \bullet \bullet'} M^{m}{}_{\bullet'}{} ^{\star^\prime}M^{n}{}_{\bullet}{}^\star+N^{(0m) \bullet \bullet'} M^{r}{}_{\bullet'}{} ^{\star^\prime}M^{n}{}_{\bullet}{}^\star+N^{(0n) \bullet \bullet'} M^{r}{}_{\bullet'}{} ^{\star^\prime}M^{m}{}_{\bullet}{}^\star
$$
$$
+{2 \over 3} N^{(rn) \bullet \bullet'} M^m{}_{\bullet'}{} ^{\star^\prime}M^0{}_{\bullet}{}^\star+{2 \over 3} N^{(0m) \bullet \bullet'} M^n{}_{\bullet'}{} ^{\star^\prime}M^r{}_{\bullet}{}^\star)
\eqno(A.14)
$$
$$
\times (\nabla_m \Lambda _{\star^\prime} \nabla_n \bar \Lambda_\star-(\Lambda \leftrightarrow \bar \Lambda))\,.
$$
Of course one can make use of equations (7) and (2) in order to recombine these terms in various ways. We note that the second line, corresponding to $C^{(1)}_{m n}$ above, evaluates to zero in the case of the Fierz-Pauli theory considered in this paper.

\medskip
{\bf References}
\medskip

\item{[1]} K.~Nguyen and P.~West, {\it Conserved asymptotic charges for any massless particle},
Int. J. Mod. Phys. A {\bf 37} (2022) no.36, 2250208 [arXiv:2208.08234 [hep-th]].

\item{[2]} T.~Regge and C.~Teitelboim,
{\it Role of Surface Integrals in the Hamiltonian Formulation of General Relativity},
Annals Phys.{\bf  88} (1974), 286.

\item{[3]} J.~Lee and R.~M.~Wald,
{\it ``Local symmetries and constraints},
J. Math. Phys. {\bf 31} (1990), 725-743.

\item{[4]} R.~M.~Wald and A.~Zoupas,
{\it A General definition of 'conserved quantities' in general relativity and other theories of gravity},
Phys. Rev. D{\bf  61} (2000), 084027
[arXiv:gr-qc/9911095 [gr-qc]].

\item{[5]} G.~Barnich and F.~Brandt,
{\it Covariant theory of asymptotic symmetries, conservation laws and central charges},
Nucl. Phys. B {\bf 633} (2002), 3-82
[arXiv:hep-th/0111246 [hep-th]].

\item{[6]} G.~Barnich, N.~Bouatta and M.~Grigoriev, {\it Surface charges and dynamical Killing tensors for higher spin gauge fields in constant curvature spaces}, JHEP {\bf 10} (2005), 010
[arXiv:hep-th/0507138 [hep-th]].

\item{[7]} G.~Barnich and G.~Compere,
{\it Surface charge algebra in gauge theories and thermodynamic integrability},
J. Math. Phys. {\bf 49} (2008), 042901
[arXiv:0708.2378 [gr-qc]].

\item{[8]} A.~Campoleoni, M.~Henneaux, S.~H\"ortner and A.~Leonard,
{\it Higher-spin charges in Hamiltonian form. I. Bose fields},
JHEP {\bf 10} (2016), 146 [arXiv:1608.04663 [hep-th]].

\item{[9]} E.~P.~Wigner,
{\it On Unitary Representations of the Inhomogeneous Lorentz Group},
Annals Math. {\bf 40} (1939), 149-204.

\item{[10]} L.~P.~S.~Singh and C.~R.~Hagen,
{\it Lagrangian formulation for arbitrary spin. 1. The boson case},
Phys. Rev. D {\bf 9} (1974), 898-909.

\item{[11]} L.~P.~S.~Singh and C.~R.~Hagen,
{\it Lagrangian formulation for arbitrary spin. 2. The fermion case},
Phys. Rev. D {\bf 9} (1974), 910-920.

\item{[12]} C.~Fronsdal,
{\it Massless Fields with Integer Spin},
Phys. Rev. D {\bf 18} (1978), 3624.

\item{[13]} J.~Fang and C.~Fronsdal,
{\it Massless Fields with Half Integral Spin},
Phys. Rev. D {\bf 18} (1978), 3630.

\item{[14]} J.~M.~F.~Labastida and T.~R.~Morris,
{\it Massless mixed symmetry bosonic free fields},
Phys. Lett. B {\bf 180} (1986), 101-106
 
\item{[15]} J.~M.~F. Labastida, {\it Massless Bosonic Free Fields},  Phys. Rev. Lett. {\bf 58} (1987)  531.

\item{[16]} J.~M.~F. Labastida, {\it Massless Particles in Arbitrary Representations of the Lorentz Group}, Nucl. Phys. B {\bf 322} (1989)  185--209.

\item{[17]} X.~Bekaert and N.~Boulanger,  {\it On geometric equations and duality for free
  higher spins}, Phys. Lett. B {\bf 561} (2003)  183--190,  [arXiv:hep-th/0301243].

\item{[18]} X.~Bekaert and N.~Boulanger,
{\it Tensor gauge fields in arbitrary representations of GL(D,R): Duality and Poincare lemma},
Commun. Math. Phys. {\bf 245} (2004), 27-67
[arXiv:hep-th/0208058 [hep-th]].

\item{[19]} X.~Bekaert and N.~Boulanger,
{\it Tensor gauge fields in arbitrary  representations of GL(D,R). II. Quadratic actions}, Commun. Math. Phys.
  {\bf 271} (2007)  723--773, [hep-th/0606198].

\end